\documentclass{aa}

\usepackage{graphicx}
\usepackage{txfonts}
\usepackage{threeparttable}
\usepackage{hyperref}

\defcitealias{Mikolaitis18}{Paper~I}
\defcitealias{Mikolaitis19}{Paper~II}
\defcitealias{Tautvaisiene20}{Paper~III}

\begin{document}

   \title{Carbon and nitrogen as indicators of stellar evolution and age}
\subtitle{A homogeneous sample of 44 open clusters from the $Gaia$-ESO Survey\thanks{Based on observations collected with the FLAMES instrument at VLT/UT2 telescope (Paranal Observatory, ESO, Chile), for the $Gaia$-ESO Large Public Spectroscopic Survey (188.B-3002, 193.B-0936).}}

   \author{G. Tautvai\v{s}ien\.{e}
   \inst{1}, 
   A. Drazdauskas
   \inst{1},
   \v{S}. Mikolaitis
   \inst{1},
   R. Minkevi\v{c}i\={u}t\.{e}
   \inst{1},   
   E. Stonkut\.{e}
   \inst{1},
   S. Randich
   \inst{2},   \\
   A. Bragaglia
   \inst{3},   
   L. Magrini
   \inst{2},   
   R. Smiljanic
   \inst{4},   
   M. Ambrosch
   \inst{1},   
   V. Bagdonas
      \inst{1}, 
   G. Casali
   \inst{3,5,6}, \\
   Y. Chorniy
   \inst{1},  
   C. Viscasillas V\'{a}zquez
    \inst{1}
          }

   \institute{Vilnius University, Faculty of Physics, Institute of Theoretical Physics and Astronomy, Saul\.{e}tekio av. 3, 10257 Vilnius, Lithuania\\
              \email{grazina.tautvaisiene@tfai.vu.lt}
        \and
INAF - Osservatorio Astrofisico di Arcetri, Largo E. Fermi, 3, 50125 Firenze, Italy
\and
INAF - Osservatorio di Astrofisica e Scienza dello Spazio, via P. Gobetti 93/3, 40129 Bologna, Italy             
             \and
Nicolaus Copernicus Astronomical Center, Polish Academy of Sciences, ul. Bartycka 18, 00-716 Warsaw, Poland
   \and
    Research School of Astronomy \& Astrophysics, Australian National University, Cotter Rd., Weston, ACT 2611, Australia
    \and
    ARC Centre of Excellence for All Sky Astrophysics in 3 Dimensions (ASTRO3D), Stromlo, Australia
             }

   \date{Received 27 May, 2025 / Accepted 28 September, 2025}

\titlerunning{Carbon and nitrogen as indicators of stellar evolution and age}

\authorrunning{G. Tautvaišienė et al.}

  \abstract
{
Low- and intermediate-mass giants undergo a complex chemical evolution that has yet to be observationally probed. The influence of core helium flash on the chemical composition of stellar atmospheres has been an open question since its theoretical prediction 60 years ago.  }
{
Based on high-resolution spectral observations of 44 open star clusters in the $Gaia$-ESO survey, our aim is to perform the first large-scale homogeneous investigation into the carbon and nitrogen photospheric content of low- and intermediate-mass giant stars in different phases of evolution. }
{
We determined carbon and nitrogen abundances using spectral synthesis of the ${\rm C}_2$ Swan (1,0) band head at 5135~{\AA} and ${\rm C}_2$ Swan (0,1) band head at 5635.5~{\AA}, $^{12}{\rm C}^{14}{\rm N}$ bands in the interval 6470 -- 6490~{\AA}, and the forbidden [O\,{\sc i}] line at 6300.31~{\AA}.}
%
{
We revealed differences in C/N abundance ratios between pre- and post-core-He-flash stars. The lower C/N ratios in core He-burning red clump stars are mainly due to the enhancement of nitrogen abundances.  
We presented calibrations of the relationship between [C/N] and stellar age for solar metallicity low- and intermediate-mass giants taking into account different evolutionary stages.}
{The C/N abundance ratios in the investigated first-ascent giant stars are slightly less affected by the first dredge-up than predicted by the theoretical models. The rotation-induced extra mixing is not as efficient as theoretically predicted. The core helium flash may trigger additional alterations in carbon and nitrogen abundances that are not yet theoretically modelled. We found that the evolutionary stage of stars must be taken into account when using [C/N] as an age indicator.  }


\maketitle

\section{Introduction}
\label{sec:intro}

Photospheric abundances of carbon and nitrogen change as a star evolves. Firstly, this happens during the first dredge-up (1DUP) at the base of the red giant branch (RGB) \citep{bib1}. During this phase, isotopes $^{13}$C and $^{14}$N diffuse outwards, while $^{12}$C diffuses inwards, which implies a decrease in the  \textsuperscript{12}C/\textsuperscript{13}C and C/N ratios. Then, during the evolution up on the RGB, stars with masses below $\sim 2.2\,M_\odot$ experience extra mixing when the hydrogen-burning shell encounters the chemical discontinuity created by the convective envelope at its maximum penetration during the 1DUP. When the hydrogen-burning shell reaches the previously mixed H-rich zone, the corresponding decrease in molecular weight of the H-burning layers induces a temporary drop in total stellar luminosity, which creates a bump in the luminosity function. During this stellar evolutionary stage the so-called extra mixing starts acting and lasts at least until the tip of the RGB. 

The exact mechanisms for deep mixing in stars are not known. Possible theoretical arguments are: rotational mixing (\citealt{Sweigart1979}; \citealt{Palacios2003}; \citealt{Chaname2005}), magnetic fields (\citealt{Hubbard1980}; \citealt{Busso2007}; \citealt{Nordhaus2008}; \citealt{Palmerini2009}; \citealt{Charbonnel2017}), internal gravity waves (\citealt{Zahn1997}), meridional circulation and turbulent diffusion (\citealt{Denissenkov2000}), thermohaline mixing (\citealt{Charbonnel2007}; \citealt{Eggleton2008}; \citealt{Cantiello2010}; \citealt{bib15}; \citealt{bib3}), combination of thermohaline mixing and magnetic fields (\citealt{Denissenkov2009, Denissenkov2024}), and combination of thermohaline mixing and rotation (\citealt{bib15}; \citealt{Lagarde2012}). In our work, we compare the C/N abundance ratios with the digitally presented models of thermohaline-induced mixing (\citealt{bib14}), and thermohaline- and rotation-induced mixing acting together (\citealt{Lagarde2012}), which have recently been identified as the probable dominating processes that govern the photospheric composition of low- and intermediate-mass giant stars. 

Helium-burning ignition in the core is another important key point in the evolution of stars. It fundamentally changes the internal structure and energy production in a star. Stars less massive than $\sim$2.2~$M_{\odot}$ experience a helium flash in their degenerate cores, while in more massive stars, the core temperatures become high enough to start helium fusion gradually in their non-degenerate cores. According to \cite{bib9}, there are some elements that may be modified in the stellar photospheres that appear as abundance anomalies.  During the He flash, the primary material mixed into and above the hydrogen shell is $^{12}$C. The other major products ($^{20}$Ne, $^{24}$Mg, $^{28}$Si, and $^{32}$S) result from the hot-$\alpha$ captures that occur during the flash. The stratification of their abundance depends on the peak temperature, with the higher atomic weight species dominating for higher temperature flashes. Other products that are generated are $^{27}$A1, $^{31}$P, $^{35}$C1, $^{36}$Ar, and, if the hydrogen shell penetrates at a reasonably high temperature, $^{14}$N. Motivated by recent observations of lithium abundances in red-clump stars (\citealt{bib4}), \cite{bib8} suggested that the excitation of internal gravity waves by vigorous turbulent convection during the helium flash may provide a physical mechanism that can induce mixing and $^7$Li production. On the other hand, to account for the widespread enhancement of lithium in red clump stars, \citet{Mori2020ApJ...901..115M, Mori2021MNRAS.503.2746M} proposed an additional energy loss mechanism associated with the neutrino magnetic moment. \cite{Li2025} suggested that Li enrichment may be caused by meridional circulation and is likely to occur during the late He flash and early stages of core He burning. They treat radial mixing caused by the meridional circulation as a diffusion process.  \cite{Mallick2025} hypothesised that the core He flash and subsequent sub-flashes may enhance Li abundances in the photospheres of red clump stars and trigger heightened chromospheric activity. It is important to understand whether core He flash can cause abundance alterations in other mixing-sensiitive elements, such as carbon and nitrogen; we therefore address this question in the present work.
 
The evolutionary changes in carbon and nitrogen abundances could serve as age indicators for low- and intermediate-mass giant stars.  Theoretically, the [C/N] versus age relations were first modelled by \cite{Salaris2015} for first dredge-up giants and later by \cite{bib14} for red-clump stars. \cite{Spina16} explored carbon abundances in Galactic field stars and derived the linear [C/Fe] versus age relation. \cite{Martig2016} investigated carbon and nitrogen as age indicators using the Apache Point Observatory Galactic Evolution Experiment (APOGEE) data and asteroseismic ages determined using $Kepler$ data; however, the derived relations are only applicable to APOGEE DR\,12 data. \cite{bib10} and \cite{bib11} used open clusters to empirically determine a linear [C/N]–log(age) relation. We also address this field of research and provide robust relations of [C/N] and age for giants at different evolutionary stages using observational data of open star clusters (OCs). 

In order to address the above-mentioned questions, we investigate photospheric carbon and nitrogen abundances in low- and intermediate-mass first-ascent giants and red-clump stars in quite a large sample of 44 OCs that ensure minimisation of the differences in internal and external parameters within the cluster, which are ideally reduced to vary in one fundamental property, mass. Under otherwise similar conditions, comparing abundance trends with turn-off mass for OCs of various ages offers valuable insight into the processes by which internal stellar evolution modifies surface chemical compositions. 

The work is structured as follows. In Section~\ref{sec2}, we describe the data sample and analysis methods. In Section~\ref{sec3}, we address the evolutionary carbon and nitrogen abundances and compare them with evolutionary models. We also investigate the influence of the He flash and present the [C/N] and age relations for stars at different evolutionary stages. We present the conclusions of this work in Section~\ref{sec4}.

\section{Data and method of analysis}\label{sec2}

For our study, we used the spectra and derived parameters and abundances of the $Gaia$-ESO survey (\citealt{bib12, bib13}) that are accessible through the ESO Science Archive Facility\footnote{http://archive.eso.org/cms/data-portal.html} and the $Gaia$-ESO Survey Data Archive. These are prepared and hosted by the Wide Field Astronomy Unit, Institute for Astronomy, University of Edinburgh, which is funded by the UK Science and Technology Facilities Council\footnote{http://ges.roe.ac.uk/pubs.html}.

\subsection{Observations}

In the $Gaia$-ESO survey, observations were accomplished with the Fiber Large Array Multi-Element Spectrograph (FLAMES) multi-fiber facility \citep{bib21}. Spectra of high-resolving power ($R\approx$ 47\,000) were obtained with Ultraviolet and Visual Echelle Spectrograph,  \citealt{bib22} (UVES). The spectra were exposed onto two CCDs, which resulted in a wavelength coverage of 4700--6840~{\AA} with a gap of about 50~{\AA} in the centre. 
The spectra were reduced with the ESO UVES pipeline and the dedicated scripts described in \citep{bib23}. Radial velocities (RV) and rotation velocities ($v\,{\rm sin}\,i$) were also determined by cross-correlating all the spectra with a sample of synthetic templates specifically derived for the $Gaia$-ESO project. Information on radial velocities was particularly useful in determining the high-probability members of stellar clusters. The typical error on RVs is about 0.4~km\,s$^{-1}$. The signal-to-noise ratio in the spectra of the observed cluster stars varied, depending on their brightness, from 20 to 190. 

The selection of clusters and target stars for the $Gaia$-ESO survey  is discussed in \cite{Bragaglia2022} and \cite{bib12}. The $Gaia$-ESO survey observed 62 clusters and data for 18 more were obtained from the ESO archive and analysed in a homogeneous way. However, spectra of giants were available not for clusters, thus a measurement of CNO abundances was carried out in 44 clusters, which we discuss here. The names of the investigated OCs and their stars along with other relevant information for this study are listed in Table~\ref{table:Results}.

\subsection{Main atmospheric parameters}

The main atmospheric parameters of the stars were determined spectroscopically in parallel by 13 nodes (groups of researchers) and later homogenised. The methodology and codes used by each node, together with the final homogenisation procedure, which included the usage of the benchmark stars (\citealt{Heiter2015, Jofre2014}), are described in detail in \citet{bib24, Hourihane2023A&A...676A.129H, Worley2024A&A...684A.148W}.  A number of constraints were imposed on the input data used in the analysis to guarantee some degree of homogeneity in the final results: the use of a common line list \citep{Heiter2021}, the use of a single set of model atmospheres \citep{bib26}, the analysis of common calibration targets, and the Solar reference abundances of \cite{bib27}, which are: $A{\rm (Fe)}_{\odot}=7.45$, $A{\rm (C)}_{\odot}=8.39$, $A{\rm (N)}_{\odot}=7.78$, and $A{\rm (O)}_{\odot}=8.66$. The nodes were free to use different computing codes, use equivalent widths or synthetic spectra, and select spectral lines from \cite{Heiter2021} for analysis.
The median of the method-to-method dispersion is 55~K, 0.13~dex, and 0.07~dex for $T_{\rm eff}$, log~$g$, and [Fe/H], respectively (\citealt{bib24}). We present the atmospheric parameters and their uncertainties in columns 4--11 of Table~\ref{table:Results}.

\subsection{Carbon and nitrogen abundances}

Carbon and nitrogen abundances were determined using spectral synthesis by the $Gaia$-ESO Vilnius node as described by \cite{bib28}. The band heads of ${\rm C}_2$ Swan (1,0) at 5135~{\AA} and of ${\rm C}_2$ Swan (0,1) at 5635.5~{\AA} were investigated to determine the abundance of carbon. The ${\rm C}_2$ bands are suitable for the investigation of carbon abundances as they give the same abundances as [C\,{\sc i}] lines, which are not sensitive to deviations in non-local thermodynamic equilibrium (NLTE) (cf. \citealt{bib29, bib30}. The interval 6470 -- 6490~{\AA}, which contains the $^{12}{\rm C}^{14}{\rm N}$ bands, was used for the nitrogen abundance analysis. It was also important to determine the oxygen abundance as the abundances of carbon and oxygen are bound by the molecular equilibrium. The forbidden [O\,{\sc i}] line at 6300.31~{\AA} was investigated. Lines of [O\,{\sc i}] are considered very good indicators of oxygen abundances. It was determined that they are not only insensitive to NLTE effects but also give similar oxygen abundance results with 3D and 1D model atmospheres (cf. \citealt{bib31, bib32}). 
To model the synthetic spectra, we used the Turbospectrum synthesis code (v12.1.1, \citealt{1998A&A...330.1109A}). We refer to \cite{bib28} for more details on the method of analysis and uncertainties. 

In the final data release of the $Gaia$-ESO Survey, for 17 member stars of NGC~2141 abundances of C, N, and O were not determined due to the distortion of the oxygen line. In this study, we decided to determine the C and N abundances for those stars by taking the averaged oxygen abundance value determined from other cluster stars.   
The C, N and O abundances and uncertainties determined are presented in columns 12--17 of Table~\ref{table:Results}.

\subsection{Membership}

The OC stars in this study passed several membership checks. The first was done before observations as described in \cite{Bragaglia2022}. Then we collected the membership probabilities from the studies by \cite{Hunt2024} and \cite{Jackson2022}. Columns 22 and 23 of Table~\ref{table:Results} are dedicated to these membership probabilities. Finally, we checked the membership for all the stars using their radial velocities and [Fe/H] determined in our study, as well as their proper motions and parallaxes taken from GaiaDR3 \citep{Gaia2016, Gaia2023, Babusiaux2023}. This was important as some stars had no previous membership determination or the membership probabilities in \cite{Hunt2024} and \cite{Jackson2022} were quite different. 

\subsection{Attribution of stars to evolutionary stages}

\begin{figure}
\includegraphics[width=\columnwidth]{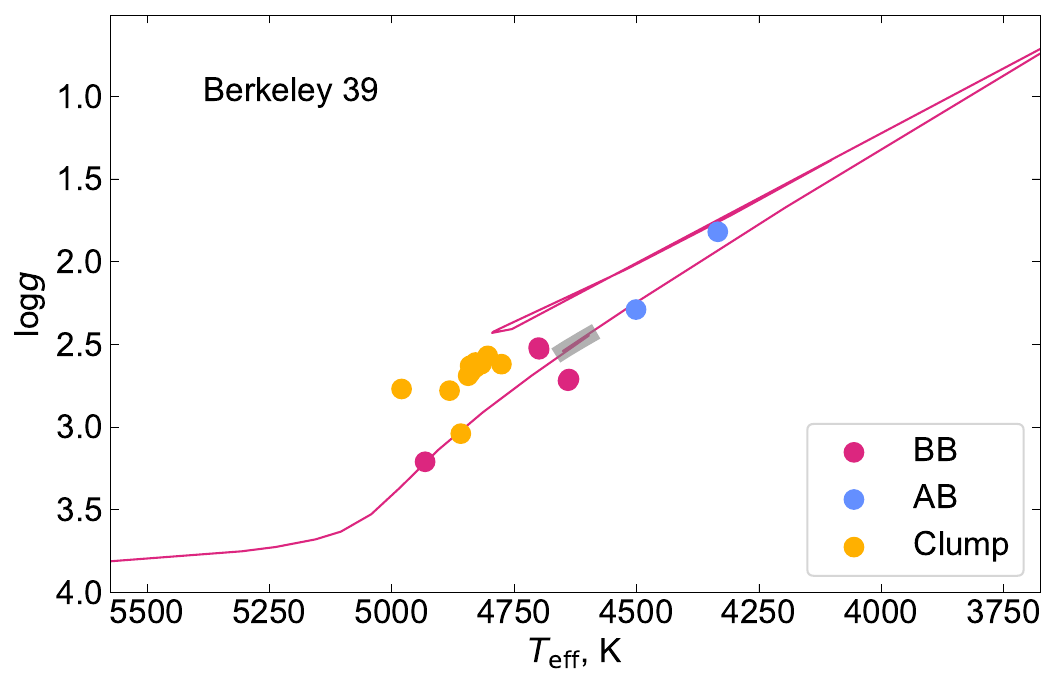}
\includegraphics[width=\columnwidth]{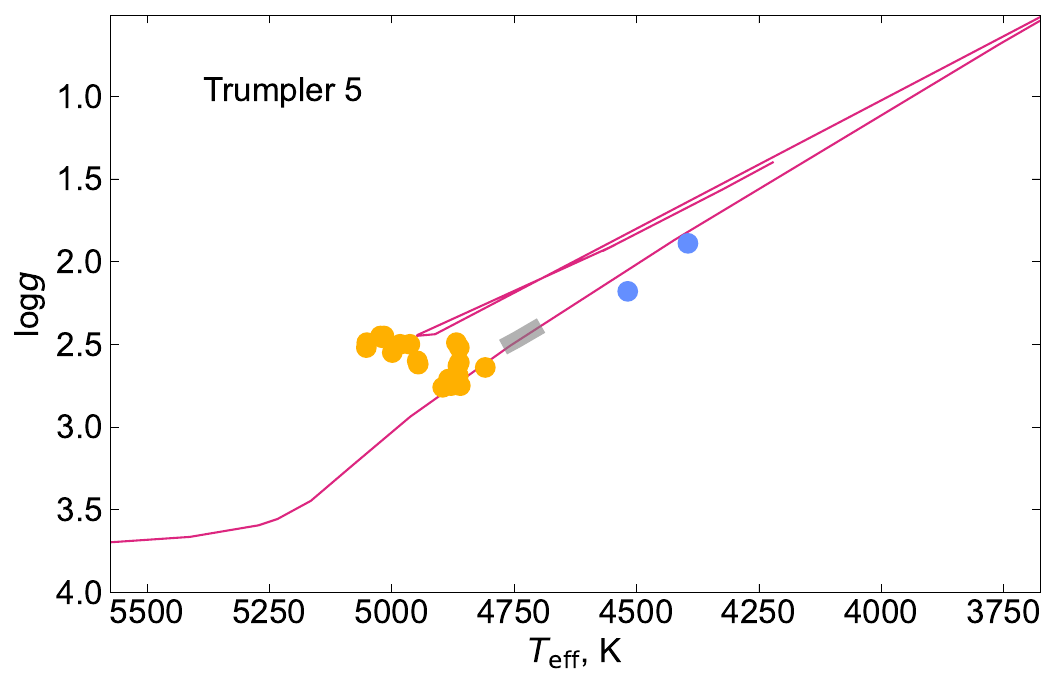}
\includegraphics[width=\columnwidth]{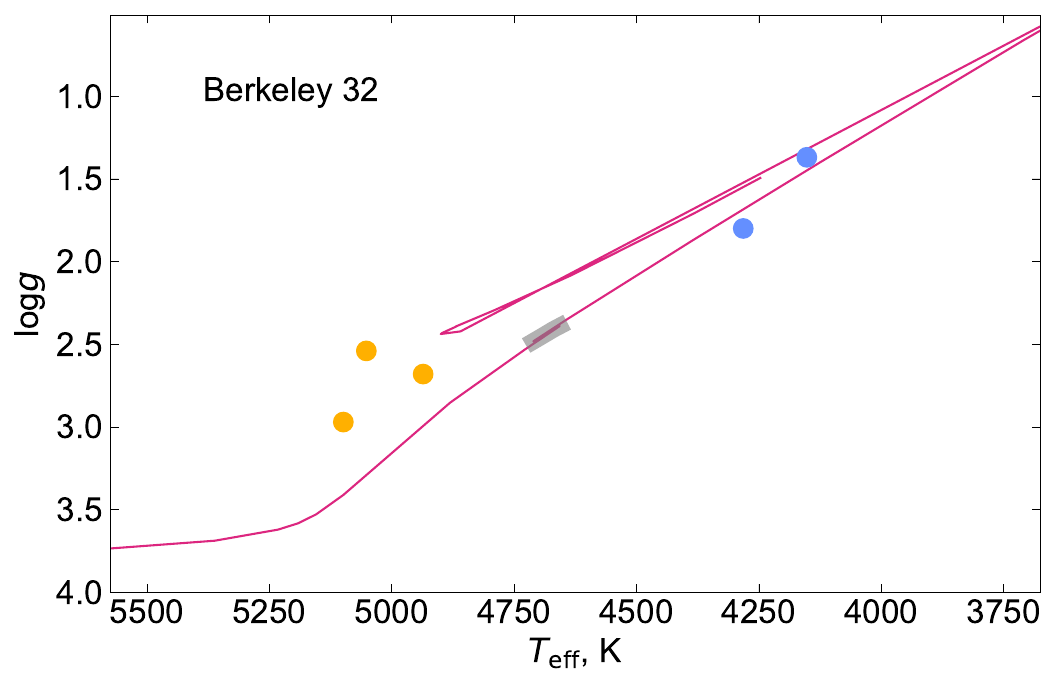}
\caption{ Positions of stars of several OCs in the log\,$g$ versus $T_{\rm eff}$ diagram together with PARSEC evolutionary sequences by \cite{bib35} corresponding to the OC age and metallicity. The RGB luminosity bump is highlighted by the grey thick stick. The red symbols mark the first ascent giants located below the red giant branch luminosity bump (BB), the blue symbols mark stars above the red giant branch luminosity bump (AB) where extra-mixing processes are acting, and the yellow symbols mark helium core burning stars (RC).}
\label{evol}
\end{figure}

In order to determine the evolutionary phase of stars, we visually investigated their positions in the log\,$g$ versus $T_{\rm eff}$ diagrams individually for each OC using the PARSEC evolutionary tracks taken from \cite{bib35}. To distinguish which stars have already passed the RGB luminosity bump, where extra mixing begins, we had to determine whether they are below or above the RGB luminosity bump (hereafter we mark them as BB and AB stars, accordingly). The RGB luminosity bump appears as a characteristic zigzag feature in stellar evolutionary tracks. For example, for a 1\,$M_\odot$ Solar-metallicity star, the luminosity bump is at about log\,$g \sim 2.55$ and $T_{\rm eff}\sim 4550$~K.  The values of the C/N ratio were also taken into account when it was difficult to separate the core-helium-burning clump stars from the neighbouring first-ascent giants (C/N ratios in OC red clump stars are lower than in RGB stars at the same log\,$g$).  As stars with turn-off masses larger than 2.2~$M_{\odot}$ do not experience extra mixing on the RGB, all of them were attributed to the BB evolutionary phase. Several examples are presented in Fig.~\ref{evol}. The attributed evolutionary phases are presented in Column 19 of Table~\ref{table:Results}.  

\subsection{Overview of the investigated OC sample}

For the analysis in this work, we have carbon and nitrogen abundances for 327 evolved stars (including C and N abundances determined for 17 stars in this work) in 44 OCs. Moreover, 18 OCs have observations of giant stars in several evolutionary stages, that is, before and after the RGB luminosity bump, or in the core helium-burning phase. In Table~\ref{table:averaged}, we present the averaged values of the [C/H], [N/H], [C/N], and C/N ratios for stars of the same evolutionary phase for every OC investigated, as well as the cluster turn-off mass, age, Galactocentric distance, [Fe/H], [O/Fe], scatter in the mean values, and numbers of stars. We used ages and Galactocentric distances of open clusters from \cite{bib19}. The turn-off mass was derived by plotting an isochrone from PARSEC with the adopted metallicity and age for each cluster, and then the turn-off point was read from the isochrone tables.

\begin{figure}
\includegraphics[width=\columnwidth]{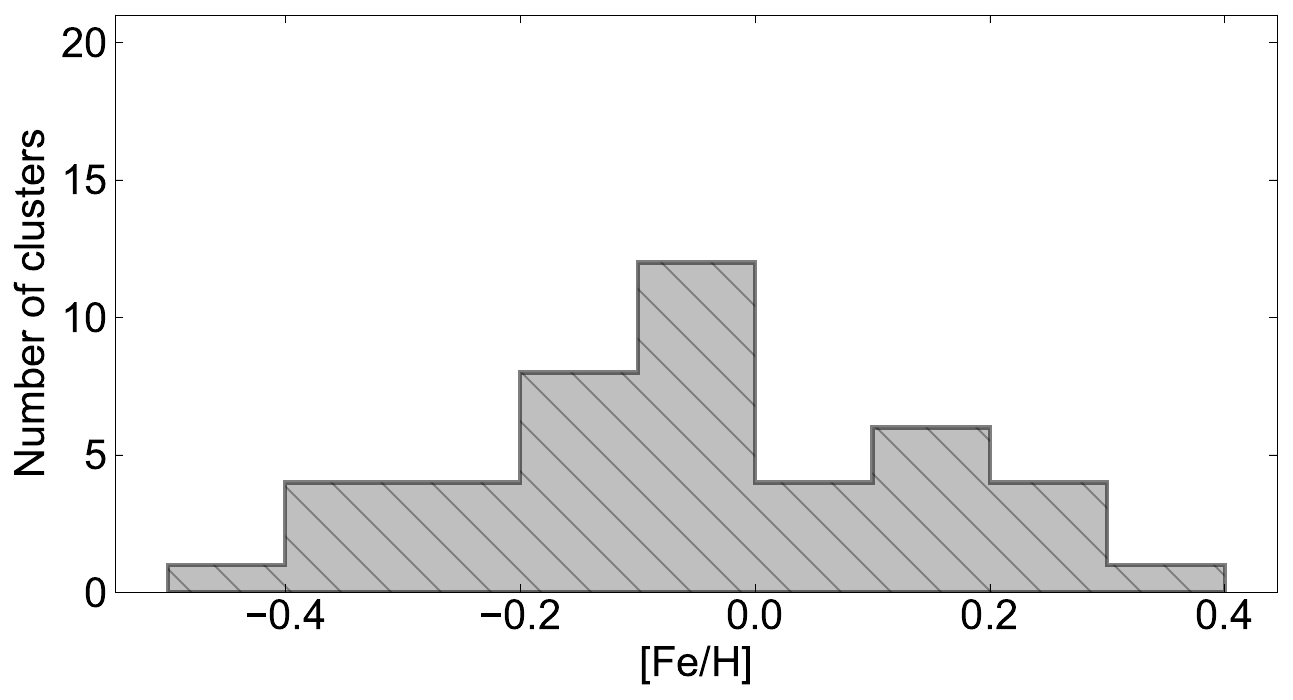}
\includegraphics[width=\columnwidth]{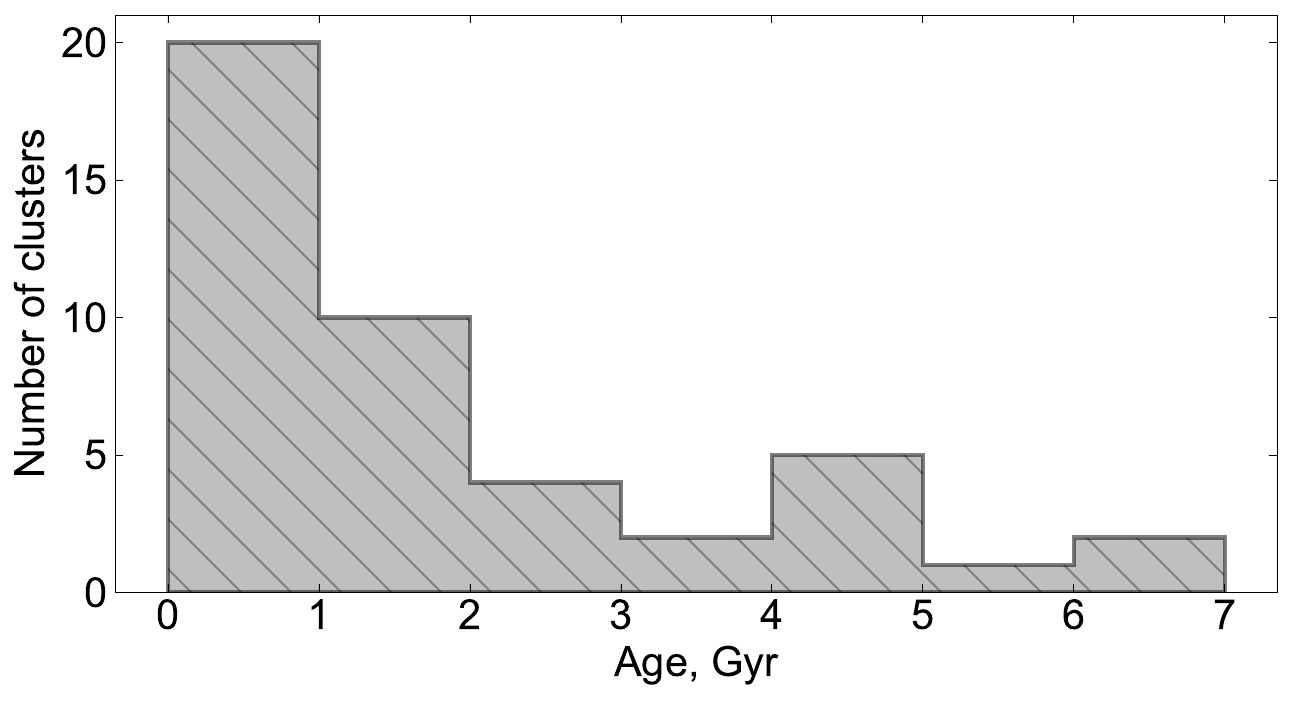}
\includegraphics[width=\columnwidth]{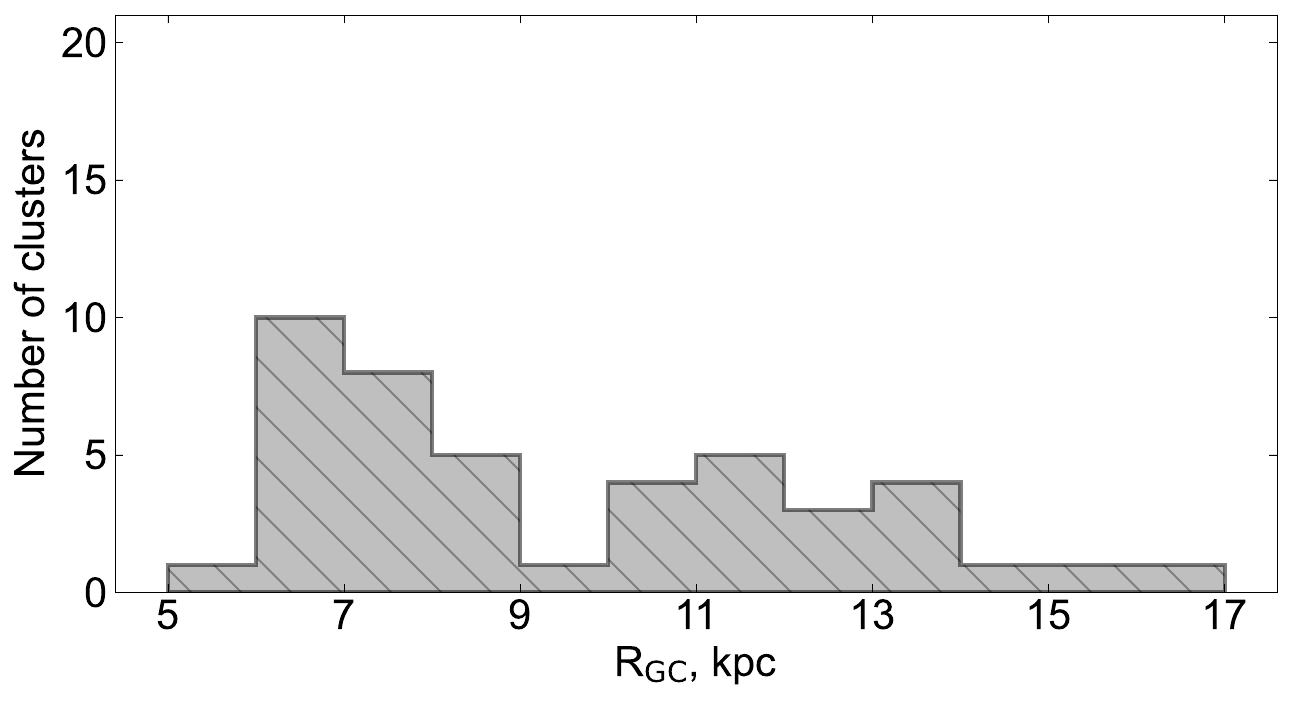}
\caption{Histograms of metallicity, age, and Galactocentric distances of the investigated open clusters with CNO abundances determined.  }
\label{histograms}
\end{figure}

The investigated OCs have rather close-to-Solar metallicities and span rather large intervals of age and Galactocentric distance. In Fig.~\ref{histograms}, we present the distributions of these parameters. 

\section{Results and discussion}\label{sec3}

\subsection{Evolutionary carbon and nitrogen abundances}

\begin{figure}
\includegraphics[width=\columnwidth]{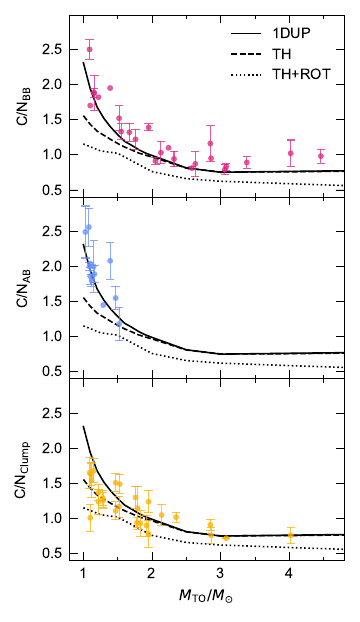}
\caption{Comparison of the averaged C/N ratios in OC stars at different  evolutionary stages with theoretical models. The solid line represents the C/N ratios predicted by the model for stars at the 1DUP and the dashed line represents the minimal values in the model with TH, both taken from \cite{bib14}. The dotted line represents the model including the thermohaline- and rotation-induced mixing acting together from  \cite{Lagarde2012}. In the upper panel, the red symbols mark the first ascent giants located below the red giant branch luminosity bump (BB). In the middle panel, the blue symbols mark stars above the red giant branch luminosity bump (AB) where extra-mixing processes are acting, and in the bottom panel, the yellow symbols mark core helium-burning stars (Clump).}
\label{fig1}
\end{figure}

Fig.~\ref{fig1} shows the averaged C/N ratios of stars in each OC according to their evolutionary stage and the model predictions for stars at the 1DUP phase and the minimal values in the model with thermohaline-induced extra mixing (TH), both taken from \cite{bib14} for Solar metallicity stars. From the top panel of Fig.~\ref{fig1}, we can see that the C/N ratios in stars between the 1DUP and the RGB luminosity bump (BB) are less affected by mixing and lie systematically $0.16\pm0.17$ (the median is 0.13) above the theoretical prediction of C/N alterations caused by the 1DUP. 

The middle panel shows stars above the RGB luminosity bump (AB), during which the extra mixing starts acting and lasts at least until the tip of the RGB. During this evolutionary phase, similarly to the 1DUP, the chemical elements $^{13}$C and $^{14}$N again diffuse outwards, while $^{12}$C diffuses inwards, which implies an additional decrease in the  \textsuperscript{12}C/\textsuperscript{13}C and  C/N ratios (\citealt{bib15, bib14}, and references therein). Here we do not see obvious deviations from stars affected by the 1DUP  (the same systematic difference of $0.15\pm0.24$ is observed with a smaller median value of 0.10), because in almost all clusters except two the stars evolved not far above the bump. This confirms the prediction that the extra mixing is not a violent event and proceeds slowly. 

The bottom panel of Fig.~\ref{fig1} shows the C/N ratios in stars that experience He burning in their cores (OC red-clump stars). Here we clearly see that the C/N ratios in stars with the lowest turn-of masses are lower than in the first-ascent giants, and are thus affected by extra mixing. The deviation from the model of the thermohaline-induced mixing for stars with turn-off masses smaller than 2.2~$M_{\odot}$ is just $0.06\pm0.19$ and the median is $0.05$. 
For stars with masses above $\sim2.2\,M_\odot$, no thermohaline-induced extra mixing is expected on the RGB (cf. \citealt{bib3}) and another so-called second dredge-up only starts on the asymptotic giant branch (e.g. \citealt{bib17}), unless rotation-induced mixing emerges \citep{bib15}. 
In these core He-burning stars, the C/N values are lower than in the 1DUP giants and the deviation from the 1DUP model is $0.05\pm0.08$ with the median value of $0.00$.
The comparison with the model (\citealt{Lagarde2012}), which includes both the thermohaline- and rotation-induced mixing, shows that the rotation-induced mixing is not as effective as theoretically modelled, stars in the entire interval of masses lie above the model by $0.26\pm 0.18$ with the same median value.

\subsection{Influence of the He flash}

\begin{figure}
\includegraphics[width=\columnwidth]{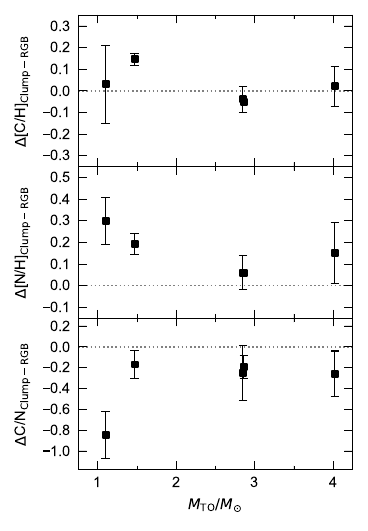}
\caption{Differences in [C/H], [N/H], and C/N ratios for post- and pre-core-He-burning stars versus turn-off mass. From left to right the data for Berkeley\,32, NGC\,2141, NGC\,6705, Berkeley\,30, and NGC\,6067 are plotted. See text for further explanations. }
\label{diff}
\end{figure}

In order to investigate the influence of He ignition in the stellar core on the surface carbon and nitrogen abundances, it is important to compare abundances in stars close to the tip of the RGB and in the red clump.    
In our sample, we have Berkeley\,32 and NGC\,2141, which have C and N abundance determinations both in stars close to the RGB tip and in the clump, and also Berkeley\,30, NGC\,6067, and NGC\,6705, which have stars with masses too large to experience thermohaline-induced mixing in their first ascent on RGB, thus the influence of the core He burning on carbon and nitrogen abundances can be examined. We find that the averaged C/N ratios of clump stars in these OCs are lowered by 0.85--0.2 compared to C/N values in pre-core-He-burning stars on the RGB. Fig.~\ref{diff} shows how these differences depend on the OC turn-off mass. The difference is larger in low-mass stars, which may experience the influence of not only the He flash but also the continuation of thermohaline-induced mixing or other processes. 

In the $Gaia$-ESO survey, the influence of He flash on lithium abundances in OCs was investigated by \cite{Magrini2021A&A...655A..23M}. 
They selected six OCs with well-populated red clumps and found that about 35\% of their clump stars have Li abundances that are similar to or higher than those of upper RGB stars. It was concluded that this can
be a sign of fresh Li production, because normally the lithium abundance decreases because of extra mixing acting on the upper RGB, and red clump stars are expected to have systematically lower surface Li abundances. 

In our sample of OCs, the lower C/N ratios in the red clump stars are mainly due to the enhancement of nitrogen abundances. 
This finding is in contrast to the study of nitrogen abundances in Galactic field stars based on the APOGEE DR12 data release  \citep{bib18}. The authors found approximately the same nitrogen abundance in RGB stars below the luminosity bump and in the clump stars, and came to the conclusion that nitrogen has been depleted between the RGB tip and the red clump during the He flash, even though there is no theoretical explanation for that yet. However, in later APOGEE data releases, DR16 and DR17, C and N abundance determinations were repeatedly reviewed and systematically shifted in order to improve them. For example, \cite{bib11} compared the C and N abundances
for stars in the 75 cluster sample that have abundances in both
DR14 and DR17. The median change for C was measured to be
$-0.029$~dex with a scatter of 0.047~dex and the median change
for N was measured to be +0.081~dex with a scatter of 0.052~dex.
Similarly, for DR16 and DR17, the median change for C
was measured to be +0.065~dex with a scatter of 0.039~dex and
the median change for N was measured to be $-0.013$~dex with a
scatter of 0.051~dex.

\begin{figure}
\includegraphics[width=\columnwidth]{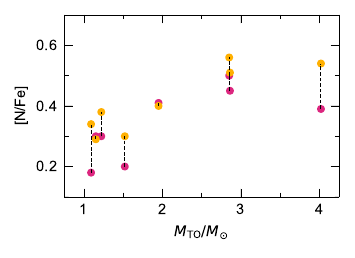}
\caption{[N/Fe] abundance ratios in 1DUP RGB stars below the luminosity bump (the red symbols) and in the red clump stars (the yellow symbols) as a function of turn-off mass in the same cluster, showing larger nitrogen abundances in the red clump stars. }
\label{[N/Fe]-RGB Clump}
\end{figure}

In Fig.\ref{[N/Fe]-RGB Clump}, we plot the averaged [N/H] abundances in 1DUP RGB stars below the luminosity bump and in the red clump of the same cluster. The abundances of [N/Fe] are not the same as was determined in \cite{Masseron2015MNRAS.453.1855M} and are larger in the red-clump stars, as expected from theory.

\subsection{Ratio of [C/N] as age indicator}

\begin{figure}
\includegraphics[width=\columnwidth]{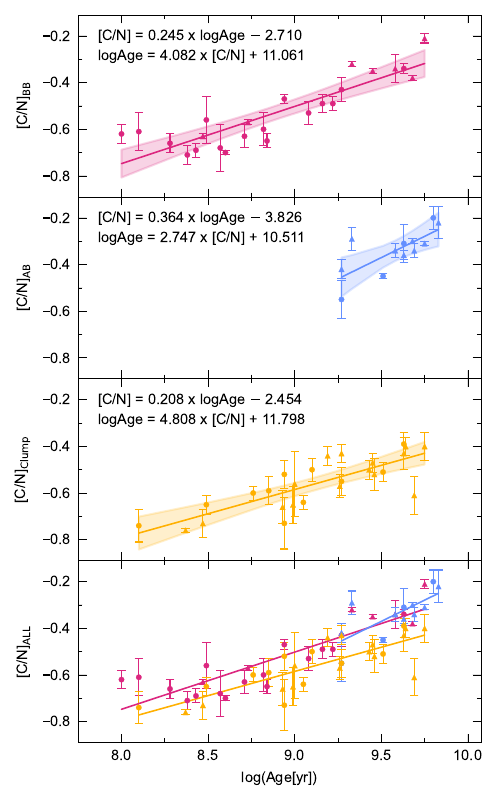}
\caption{Relations of [C/N] with age for OC stars at different evolutionary stages. The shadowed areas show the confidence intervals of 95\%. The meaning of the symbols is as in Fig.~\ref{fig1}, only the clusters with $R_{\rm GC}\le 9.5$~kpc are marked by triangles, and those with $R_{\rm GC}> 9.5$~kpc are marked by circles.}
\label{fig2}
\end{figure}

\cite{bib10} presented a relation between the C/N ratio in evolved stars and their age based on preliminary data for 17 and 23 open clusters from the $Gaia$-ESO and APOGEE DR14  surveys, respectively. Several clusters had data in both studies. However, in the APOGEE DR17 data release, the C and N abundance determinations were recalibrated and systematically shifted. Thus, since we have the final $Gaia$-ESO Survey data for stars in 44 open clusters with determined C and N abundances, we can constrain the [C/N] versus age relation in a more homogeneous way. 
As changes in photospheric abundances of carbon and nitrogen increase as stars evolve, we investigated the individual [C/N] and age relations for stars very carefully divided into three evolutionary phases (below and above the red giant branch luminosity bump and in the clump).

To determine the linear regression fit of our data points, we employed a bootstrapping method with 10\,000 iterations, where we slightly varied the values of [C/N] with each calculation. The [C/N] values were varied considering the scatter in the mean [C/N] values and drawing a value randomly from a Gaussian distribution for each point at each time. Then, the final values for slope and intercept were calculated as averages of the 10\,000 linear regression fits that had been calculated.

We also calculated the confidence intervals around our linear regression model. We used a critical value of 0.975 to calculate the confidence intervals of 95\%. With this in mind, they should accurately reflect where the linear regression model lies. In addition, the average Pearson correlation coefficients are 0.84, 0.67, and 0.71 for the samples BB, AB, and Clump, respectively. We also performed a permutation test, randomly reassigning up to 10\% of the objects between groups. The results remained within 1~sigma of the reported values, with the different trends persisting for the different groups.

The obtained [C/N] and age relations as presented in Fig.~\ref{fig2} are the following:

\begin{itemize}

    \item
RGB stars below the RGB luminosity bump \\
log\,Age = 4.082 x [C/N] + 11.061, PPC=0.84 
\\
\item
RGB stars above the RGB luminosity bump \\
log\,Age = 2.747 x [C/N] + 10.511, PPC=0.67
\\
\item
Core He-burning stars\\
log\,Age = 4.808 x [C/N] + 11.798, PPC=0.71 
\end{itemize}

There is no obvious deviation in the relations for stars in the evolutionary phases below and above the RGB luminosity bump because in the sample clusters stars evolved not far above the bump. It is interesting to see that \cite{Roberts2024} also do not find compelling evidence of extra mixing in the [C/N] ratios for the close-to-Solar metallicity first-ascent giants in the evolutionary phase above the RGB luminosity bump. Probably in their sample, as in ours, the majority of those stars evolved not far away from the RGB luminosity bump. 

As the sample of clusters has two peaks in the $R_{\rm GC}$ distribution separated at about 9.5~kpc, we decided to mark those subsamples of stars with different symbols in Fig.~\ref{fig2} (the clusters with $R_{\rm GC}\le 9.5$~kpc are marked by triangles, and those with $R_{\rm GC}> 9.5$~kpc are marked by circles) and see if there are differences related to the Galactocentric distance. Several AB stars lie slightly above the relation. This may be due to their larger Galactocentric distances, where the initial CNO abundances could be different according to investigations of the cepheids {\citealt{Luck2011}, \citealt{Luck2018}) or H\,{\sc{ii}} regions} (\citealt{Arellano2020}). However, in a study of 1475 pre-dredge-up giants from APOGEE DR12 by \cite{Martig2016}, no difference in [C/N] values was found in the inner and outer Galactic discs. The number of OCs on large Galactocentric distances in our sample is not large enough to draw confident conclusions. 

\begin{figure}
\includegraphics[width=\columnwidth]{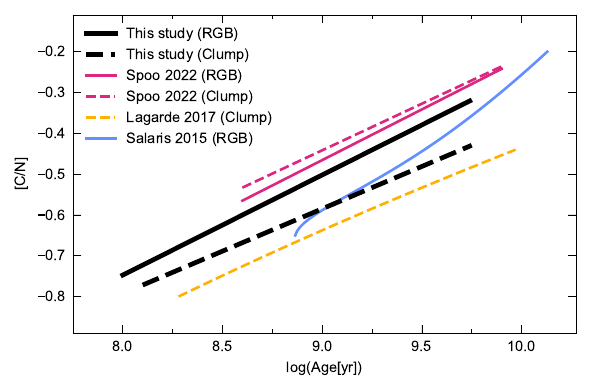}
\caption{Comparison of the [C/N]-age relations for the 1DUP RGB (the continuous lines) and red clump stars (the dashed lines). The black lines show relations derived in our work,  the red lines are for \cite{bib11}, the blue line is for \cite{Salaris2015}, and the yellow line is for \cite{bib14}.     }
\label{All studies}
\end{figure}

In Fig.~\ref{All studies}, we display the [C/N]-age relations for the 1DUP RGB and red clump stars derived in this work along with the relations from other studies. As the age calculation formula in \cite{Salaris2015} needs the [Fe/H] value as input, we used the Solar metallicity. From \cite{bib14}, we took the formula presented for stars with $-0.05<{\rm [Fe/H]}<0.05$. From \cite{bib11}, we read the [C/N]-age relations from their plots because the formula was computed and presented for the age-[C/N] relation.  
For the 1DUP RGB stars, the agreement between our work and \cite{bib11} is rather good; only the red clump stars in that study behave quite strange. The theoretical relation for the red clump stars from \cite{bib14} and for the 1DUP RGB stars from \cite{Salaris2015} would give slightly older ages for the corresponding stars.

Thus, our study shows that while using the [C/N] ratios for the age evaluation of red clump and RGB stars, individual relations have to be used. The relations are applicable for stars in the metallicity interval $-0.4 < \rm{[Fe/H]} < 0.3$ with approximate turn-off masses $1 < M/M_{\odot} < 4$ and ages from 100~Myr to 7~Gyr,  corresponding to the sample of OCs investigated. For more metal-deficient stars, the empirical relations are more challenging and have yet to be investigated.

\section{Conclusions}\label{sec4}

Unprecedentedly precise high-resolution spectral analysis of 327 low- and intermediate-mass evolved stars in 44 open clusters by the $Gaia$-ESO survey allowed us to draw the following conclusions: 

\begin{itemize}
\item
The C/N abundance ratios in the investigated first-ascent RGB stars are affected by the first dredge-up slightly less than predicted by the theoretical model of \cite{bib14}. The rotation-induced extra mixing is not as efficient as predicted by the model of \cite{Lagarde2012}.  

\item
Based on a sample of five open clusters with both pre- and post-core-He-burning-ignition stars investigated, the influence of the core He ignition on carbon and nitrogen abundances was revealed. We found that the averaged C/N ratios of clump stars in these OCs are lowered by 0.85--0.2 compared to C/N values in pre-core-He-burning RGB stars. The difference is larger in low-mass stars, which may experience not only processes triggered by the core He flash but also the continuation of thermohaline-induced mixing. The lower C/N ratios are mainly due to the enhancement of nitrogen abundances.

\item 
When using [C/N] abundance ratios as stellar age indicators for the first-ascent giant stars and core-helium-burning stars, separate relations have to be used. In this work, we provide such relations applicable to giant stars in the metallicity interval $-0.4 < \rm{[Fe/H]} < 0.3$ with approximate turn-off masses $1 < M/M_{\odot} < 4$ and ages from 100~Myr to 7~Gyr. 

\end{itemize}

\section*{Data availability}

Tables \ref{table:Results} and \ref{table:averaged} are only available in electronic form at the CDS via anonymous ftp to cdsarc.u-strasbg.fr (130.79.128.5) or via http://cdsweb.u-strasbg.fr/cgi-bin/qcat?J/A+A/.

\begin{acknowledgements}

We thank the referee for suggestions that improved the presentation and clarity of the article.
Based on data products from observations made with ESO Telescopes at the La Silla Paranal Observatory under programmes ID 188.B-3002, 193-B-0936, and 197.B-1074. These data products have been processed by the Cambridge Astronomy Survey Unit (CASU) at the Institute of Astronomy, University of Cambridge, and by the FLAMES/UVES reduction team at INAF-Osservatorio Astrofisico di Arcetri. Public access to the data products is via the ESO Archive, and the Gaia-ESO Survey Data Archive, prepared and hosted by the Wide Field Astronomy Unit, Institute for Astronomy, University of Edinburgh, which is funded by the UK Science and Technology Facilities Council. 
This work has made use of data from the European Space Agency (ESA) mission {\it Gaia} (\url{https://www.cosmos.esa.int/gaia}), processed by the {\it Gaia} Data Processing and Analysis Consortium (DPAC,
\url{https://www.cosmos.esa.int/web/gaia/dpac/consortium}). Funding for the DPAC has been provided by national institutions, in particular the institutions participating in the {\it Gaia} Multilateral Agreement.
We have made extensive use of the NASA ADS and SIMBAD databases.
G.T., A.D., \v{S}.M., R.M., E.S., M.A., V.B., Y.Ch., and C.V.V. acknowledge funding from the Research Council of Lithuania (LMTLT, grant No. S-MIP-23-24).  L.M. thanks INAF for the support (Large Grants EPOCH and WST), the Mini-Grants Checs (1.05.23.04.02), and the financial support under the National Recovery and Resilience Plan (NRRP), Mission 4, Component 2, Investment 1.1, Call for tender No. 104 published on 2.2.2022 by the Italian Ministry of the University and Research, funded by the European Union – NextGenerationEU – Project ‘Cosmic POT’ Grant Assignment Decree No. 2022X4TM3H by the Italian Ministry of the University and Research. G.C. thanks the Mini-Grants Checs (1.05.23.04.02).
\end{acknowledgements}

\bibliographystyle{aa} 

\bibliography{aa55685-25corr}        

\appendix

\onecolumn

\section{Machine readable tables of results}

\begin{table}[h]
\caption{Atmospheric parameters, CNO abundances, membership probabilities, and evolutionary stages of stars.}
\begin{tiny}
 \begin{tabular}{rlll}
 \hline
 \hline
 Col. & Label & Units & Explanations\\
 \hline
 1      & Cluster        & --         & Cluster name\\
 2      & ID         & --         & Identification of the star in the cluster\\
 3      & SNR             & --  & Signal-to-noise ratio \\ 
 4      & $T_{\rm eff}$     & K             & Effective temperature\\
 5      & $e$\_$T_{\rm eff}$  & K        & Uncertainty in effective temperature\\
 6      & log\,$g$               & dex & Stellar surface gravity\\
 7      & $e$\_log\,$g$            & dex & Uncertainty in stellar surface gravity\\
 8      & [Fe/H]             & dex          & Metallicity\\
 9      & $e$\_[Fe/H]          & dex        & Uncertainty in [Fe/H]\\
 10     & $V_{\rm t}$         &  km\,s$^{-1}$   & Microturbulence velocity\\
 11     & $e\_Vt$              & km\,s$^{-1}$  & Uncertainty in microturbulence velocity\\
 12     & [C/H]                  & dex              & Carbon abundance\\
 13     & $e$\_[C/H]         & dex          & Uncertainty in carbon abundance\\
 14     & [N/H]                  & dex              & Nitrogen abundance\\
 15     & $e$\_[N/H]         & dex          & Uncertainty in nitrogen abundance\\
 16     & [O/H]                  & dex              & Oxygen abundance\\
 17     & $e$\_[O/H]         & dex          & Uncertainty in oxygen abundance\\
 18     & C/N    & --  & C/N abundance ratio \\
 19     & Evol  & -- & Evolutionary stage: BB -- below the luminosity bump of RGB, AB -- above the luminosity bump of RGB, C -- Clump star \\
 20     & RV & km\,s$^{-1}$   & Radial velocity according to the $Gaia$-ESO Survey \\
 21   & $e$\_RV    & km\,s$^{-1}$  & Uncertainty in radial velocity \\
 22     & M$_{\rm HR}$      & \%     & Membership probability according to \cite{Hunt2024} \\
 23     & M$_{\rm J}$   & \%         & Membership probability according to \cite{Jackson2022} \\
 \hline
 \end{tabular}
 \tablefoot{The full version is available at the CDS.}
 \end{tiny}
 \label{table:Results}\\
 \end{table}
 \centering

 \begin{table}
 \caption{Averaged CNO abundances in stars of different evolutionary stages.}
  \begin{tiny}
  \begin{tabular}{rlll}
\hline
 \hline
 Col. & Label & Units & Explanations\\
 \hline
 1      & Cluster        & --         & Cluster name\\
 2      & [Fe/H]         & dex         & Metallicity \\
 3      & TO mass             & $M_\odot$  & Turn-off mass \\ 
 4      & Age     & Gyr            & Age\\
 5      & log\,age  &        -- & log\,age\\
 6      & $R_{\rm GC}$  & kpc  & Galactocentric distance \\
7      & $z$  & kpc & Maximal distance from the galactic plane \\
8      & [O/H]    &  dex  & Mean oxygen abundance \\
9      &  s\_[O/H]         & dex   & Scatter in the mean oxygen abundance \\
10    & [C/H]\_BB                  & dex              & Mean carbon abundance in stars below the RGB luminosity bump\\
 11     & s\_[C/H]\_BB         & dex          & Scatter of the mean carbon abundance in stars below the RGB luminosity bump\\
 12     & [N/H]\_BB                  & dex              & Mean nitrogen abundance in stars below the RGB luminosity bump\\
 13     & s\_[N/H]\_BB         & dex          & Scatter of the mean nitrogen abundance in stars below the RGB luminosity bump\\
14   & [C/N]\_BB  & dex & Mean [C/N] abundance ratio in stars below the RGB luminosity bump\\
 15     & s\_[C/N]\_BB         & dex          & Scatter of the mean [C/N] ratio in stars below the RGB luminosity bump\\
 16     & C/N\_BB    & --  & Mean C/N abundance ratio in stars below the RGB luminosity bump \\
17     & s\_C/N\_BB         & --          & Scatter of the mean C/N ratio in stars below the RGB luminosity bump\\
 18   & n\_BB  & -- & Number of stars below the RGB luminosity bump \\
19     & [C/H]\_AB                  & dex              & Mean carbon abundance in stars above the RGB luminosity bump\\
 20     & s\_[C/H]\_AB         & dex          & Scatter of the mean carbon abundance in stars above the RGB luminosity bump\\
21     & [N/H]\_AB                  & dex              & Mean nitrogen abundance in stars above the RGB luminosity bump\\
 22     & s\_[N/H]\_AB         & dex          & Scatter of the mean nitrogen abundance in stars above the RGB luminosity bump\\
 23   & [C/N]\_AB  & dex & Mean [C/N] abundance ratio in stars above the RGB luminosity bump\\
  24     & s\_[C/N]\_AB         & dex          & Scatter of the mean [C/N] ratio in stars above the RGB luminosity bump\\
 25     & C/N\_AB    & --  & Mean C/N abundance ratio in stars above the RGB luminosity bump \\
  26     & s\_C/N\_AB         & --          & Scatter of the mean C/N ratio in stars above the RGB luminosity bump\\
 27   & n\_AB  & -- & Number of stars above the RGB luminosity bump \\
 28     & [C/H]\_C                  & dex              & Mean carbon abundance in red clump stars \\
 29     & s\_[C/H]\_C         & dex          & Scatter of the mean carbon abundance in red clump stars \\
 30     & [N/H]\_C                  & dex              & Mean nitrogen abundance in red clump stars \\
 31     & s\_[N/H]\_C         & dex          & Scatter of the mean nitrogen abundance in red clump stars \\
 32   & [C/N]\_C  & dex & Mean [C/N] abundance ratio in red clump stars \\
  33     & s\_[C/N]\_C         & dex          & Scatter of the mean [C/N] ratio in red clump stars \\
 34     & C/N\_C    & --  & Mean C/N abundance ratio in red clump stars  \\
  35     & s\_C/N\_C         & --          & Scatter of the mean C/N ratio in red clump stars \\
36   & n\_C  & -- & Number of red clump stars \\
  \hline
  \end{tabular}
  \tablefoot{The full version is available at the CDS.}
    \end{tiny}
  \label{table:averaged}\\
 \end{table}
 \centering
\end{document}